\documentclass[preprint,showpacs,amsmath,amssymb,showkeys,superscriptaddress,endfloats]{revtex4}

\usepackage[latin1]{inputenc}
\usepackage{graphicx}

\usepackage{amsmath}
\usepackage{bm}
\usepackage{amssymb}
\usepackage{ae,aeguill,aecompl}
\usepackage{amstext}
\usepackage{latexsym}
\usepackage{amsfonts}
\usepackage{mathrsfs}
\usepackage{latexsym}
\usepackage{dcolumn}
\usepackage{url}
\usepackage{color}

\begin{document}

\title{Electronic temperatures, densities and plasma X-ray emission of a
14.5~GHz Electron-Cyclotron Resonance Ion Source}

\author{A.~Gumberidze\footnote{Present address: ExtreMe Matter Institute EMMI, GSI, 64291 Darmstadt, Germany}}
\affiliation{Laboratoire Kastler Brossel,
École Normale Supérieure; CNRS; Université Pierre et Marie Curie - Paris 6
Case 74; 4, place Jussieu, 75252 Paris CEDEX 05, France
}
\author{M.~Trassinelli}
\email{martino.trassinelli@insp.jussieu.fr}
\affiliation{Institut de NanoSciences de Paris; CNRS; Université Pierre et Marie Curie - Paris 6; Campus Boucicaut, 140 rue de Lourmel, Paris, 75015 France
}
\author{N.~Adrouche}
\affiliation{Institut de NanoSciences de Paris; CNRS; Université Pierre et Marie Curie - Paris 6; Campus Boucicaut, 140 rue de Lourmel, Paris, 75015 France
}
\author{C.I.~Szabo\footnote{Present address: Naval Research Laboratory, Space Science Division, 4555 Overlook Avenue SW, Washington, DC 20375, USA}}
\affiliation{Laboratoire Kastler Brossel,
École Normale Supérieure; CNRS; Université Pierre et Marie Curie - Paris 6
Case 74; 4, place Jussieu, 75252 Paris CEDEX 05, France
}
\author{P.~Indelicato}
\affiliation{Laboratoire Kastler Brossel,
École Normale Supérieure; CNRS; Université Pierre et Marie Curie - Paris 6
Case 74; 4, place Jussieu, 75252 Paris CEDEX 05, France
}
\author{F.~Haranger\footnote{Present address: Schlumberger, SRPC Clamart, 1 Rue Henri Becquerel, 92143 Clamart,France}}
\affiliation{Institut de NanoSciences de Paris; CNRS; Université Pierre et Marie Curie - Paris 6; Campus Boucicaut, 140 rue de Lourmel, Paris, 75015 France
}
\author{J.-M.~Isac}
\affiliation{Laboratoire Kastler Brossel,
École Normale Supérieure; CNRS; Université Pierre et Marie Curie - Paris 6
Case 74; 4, place Jussieu, 75252 Paris CEDEX 05, France
}
\author{E.~Lamour}
\affiliation{Institut de NanoSciences de Paris; CNRS; Université Pierre et Marie Curie - Paris 6; Campus Boucicaut, 140 rue de Lourmel, Paris, 75015 France
}
\author{E.-O.~Le~Bigot}
\affiliation{Laboratoire Kastler Brossel,
École Normale Supérieure; CNRS; Université Pierre et Marie Curie - Paris 6
Case 74; 4, place Jussieu, 75252 Paris CEDEX 05, France
}
\author{J.~Mérot}
\affiliation{Institut de NanoSciences de Paris; CNRS; Université Pierre et Marie Curie - Paris 6; Campus Boucicaut, 140 rue de Lourmel, Paris, 75015 France
}
\author{C.~Prigent}
\affiliation{Institut de NanoSciences de Paris; CNRS; Université Pierre et Marie Curie - Paris 6; Campus Boucicaut, 140 rue de Lourmel, Paris, 75015 France
}
\author{J.-P.~Rozet}
\affiliation{Institut de NanoSciences de Paris; CNRS; Université Pierre et Marie Curie - Paris 6; Campus Boucicaut, 140 rue de Lourmel, Paris, 75015 France
}
\author{D.~Vernhet}
\affiliation{Institut de NanoSciences de Paris; CNRS; Université Pierre et Marie Curie - Paris 6; Campus Boucicaut, 140 rue de Lourmel, Paris, 75015 France
}

\date{\today}

\begin{abstract}
We have performed a systematic study of the Bremsstrahlung emission from the electrons in the plasma of a commercial 14.5 GHz Electron-Cyclotron Resonance Ion Source. The electronic spectral temperature and the product of ionic and electronic densities of the plasma are measured by analyzing the Bremsstrahlung spectra recorded for several rare gases (Ar, Kr, Xe) as a function of the injected power. Within our uncertainty, we find an average temperature of $\approx 48$~keV above 100W, with a weak dependency on the injected power and gas composition. Charge state distributions of extracted ion beams have been determined as well, providing a way to disentangle  the ionic density from the  electronic density. Moreover X-ray emission from highly charged argon ions in the plasma has been observed with a high-resolution mosaic crystal spectrometer, demonstrating the feasibility for high-precision measurements of transition energies of highly charged ions, in particular of the magnetic dipole (M1) transition of He-like of argon ions.
\end{abstract}
%

\maketitle

\section{Introduction}
\label{sec:intro} 
Electron-Cyclotron Resonance Ion Sources (ECRIS) are in wide use, providing medium to highly charged ion beams, for injecting heavy ions into accelerators or to  study the interaction dynamics of ions with matter at low energy. ECRIS are made of a magnetic bottle and a hexapole that trap electrons along the beam axis (minimum-B structure) and in the perpendicular direction,respectively. Electrons are heated by injecting a high-frequency electromagnetic wave (generally from 2.2 to 28 GHz) that is resonant with the electron cyclotron frequency in the magnetic field (for more details on ECRIS, see, for example, Ref.~\cite{gel1996}).
The absorption of energy of the radio-frequency wave happens on a constant $|\boldsymbol{B}|$ surface in the source such that $2\pi f = e |\boldsymbol{B}| /m_e$ where $f$ is the
electromagnetic wave frequency and $e$ and $m_e$ are the electron charge and mass, respectively. The heated electrons ionize, by inelastic collision, the gas injected in the magnetic structure and provide ions that are trapped in the space charge of the electrons. Plasma is thus created, with ionic temperatures in the range of a few eV and electronic temperatures that can reach several tens of keV. Ions are extracted from the plasma using specific electrodes polarized to negative high-
voltages (usually from 10 to 20~ kV). Plasmas with such high energy electrons will also lead to the production of electromagnetic radiations with  
 energy up to hard X rays and that have been used for plasma diagnostics for a long time \cite{hut1987}.

More specifically, interest in studying ECR plasma through radiation in the X-ray range domain has recently risen. The electron Bremsstrahlung spectra have been used for measuring electronic temperature in the ECRIS plasma \cite{blbg1994,fkzz1996,lpcb1996,ghbb1998,kki1998,lltv2006,zzmw2008,lblt2008}, and 
high-resolution X-ray spectroscopy has been applied to characterize charge state distribution and electronic density \cite{lzz1996,gklu1998,dkgb2000,mmcs2009}. Results on accurate X-ray spectroscopy, from Ref.~\cite{dkgb2000}, combined with Multiconfiguration Dirac-Fock calculations, have led to understand the mechanisms involved in  ion formation in the ECRIS plasma \cite{cmps2001,mcsi2001}.
In addition, the high-intensity
X rays emitted by the highly charged ions in an ECRIS have been
proposed as X-ray energy standards \cite{agis2003} together with
exotic atoms.

In 2003, we have acquired a commercial 14.5~GHz , all permanent-magnets, SUPERNANOGAN \cite{bbck2000} ECR ion source, we named SIMPA (Source d'Ions Multichargés de Paris). The high-frequency should allow for efficient operations, but the use of permanent magnets leads to a reduced plasma chamber volume (in particular in diameter), which could impact performances.

In this paper, we present a systematic study of the temperature and densities in plasmas of argon, krypton and xenon, combining X-ray spectroscopy with extracted ion-beam intensity measurement. Systematic studies as a function of power were realized leading to an extensive data set of electronic temperatures as well as ionic and electronic densities of the plasma source. 

The paper is organized as follows. In Sec.~\ref{sec:setup},  the experimental setup is described in detail. Section \ref{sec:plasma} is devoted to the different measurements performed in order to extract the electronic temperatures and the product of electronic and ionic densities exploiting electron Bremsstrahlung spectra. Combined with measurements of the extracted ion currents, we present the method used to determine the mean charge state of ions in the plasma, and how electronic and ionic density can be estimated separately. High-resolution X-ray spectra of highly charged argon ions from the ECR plasma are discussed in Sec.~\ref{sec:xray} and our concluding remarks are presented in Section \ref{sec:concl}.
\section{Experimental setup}
\label{sec:setup}

\subsection{The ECRIS and the beam line}
\label{sec:beam-line} 
Details on the principle of our permanent
magnets ECRIS can be found in \cite{bbck2000}. Some indications on
typical values of magnetic fields that should be attained at
different points of a permanent ECRIS magnet at the minimum
$|\boldsymbol{B}|$, or at ejection are given in Ref.\
~\cite{hgmc2003}.
Using a Hall-probe magnetometer, the longitudinal
field in SIMPA was found to be 1.18~T for the maximum field at
injection and $B_{\textrm{min}}=0.47~$T. This leads to an injection
mirror ratio of 2.5. The field value for which the electron cyclotron
frequency is equal to the injected frequency is $B_{14.5} =
0.518$~T. The extraction field is estimated to be $\approx 0,97$~T
(the exact location of the maximum could not be reached for
mechanical reasons). From the magnetic field and 
resonance frequency measurements we deduce a plasma length of $34± 1$~mm.

\begin{figure*}[htbp]
\centering
\includegraphics[width=\textwidth,angle=0]{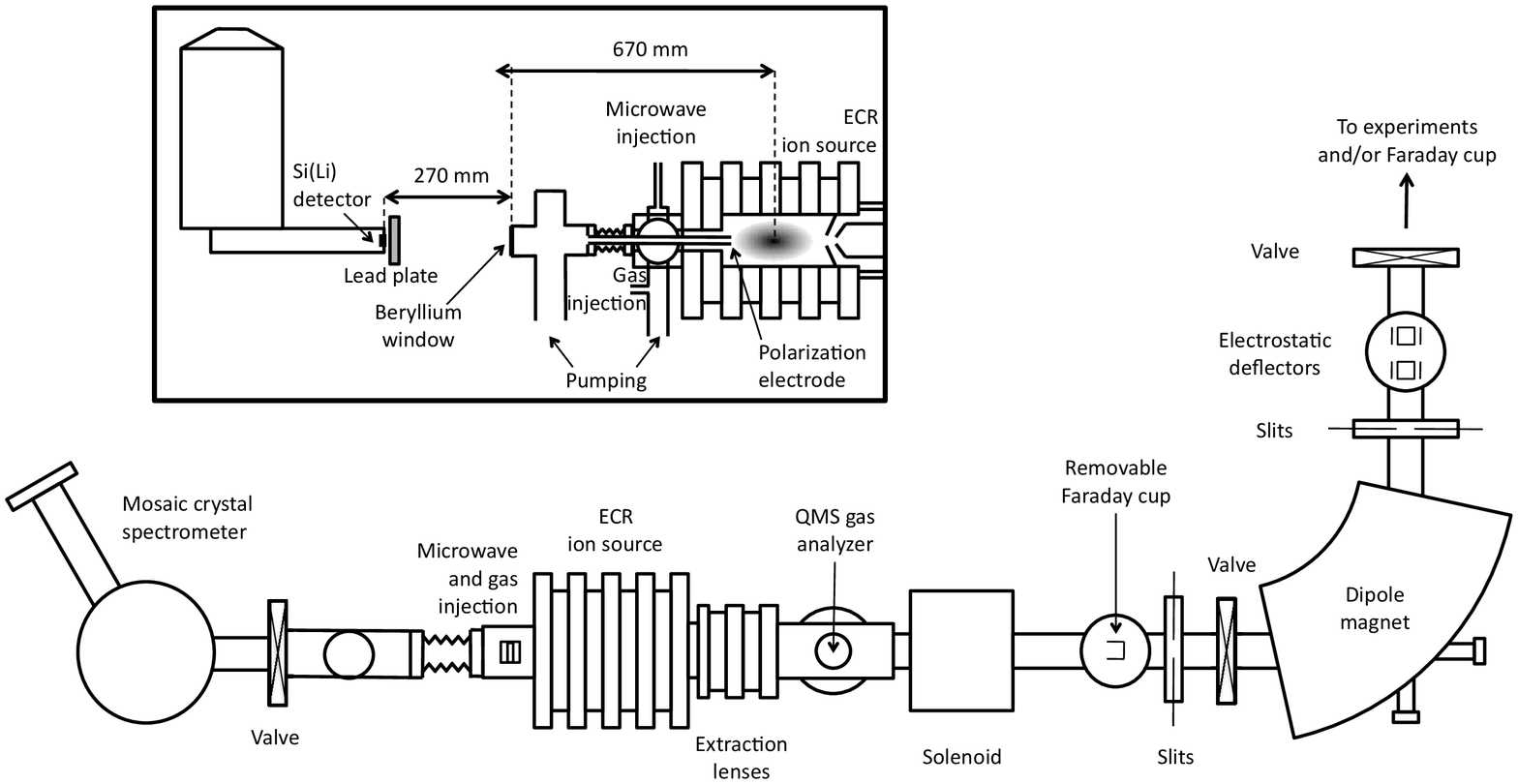}

\caption[]{Global view of the SIMPA installation including the ECR ion source and the extraction and analysis line. The detailed set-up for Bremsstrahlung measurements is displayed in the inset.
}
\label{fig:simpa-setup}
\end{figure*}

The SIMPA ECRIS microwave injection is of the original perpendicular
design, with a copper cube to which are attached the waveguide, a tuner,
and a turbo-molecular
vacuum pump, all perpendicular to the source axis.  The support and
main gas are injected by two thermo-valves and two stainless-steel
capillaries above the pump.  The tuner is composed of a piston moving
inside a tube, with a microwave-tight seal. It can be moved by a
stepping motor to optimize the  source performances. A copper
tube, centered on the source axis, and traversing completely the
copper cube, is used to deflect the microwaves into the plasma
chamber. A polarization electrode (see Fig.~\ref{fig:simpa-setup}), made of a smaller diameter copper
tube penetrates into the plasma
chamber, up to the edge of the plasma.  The plasma chamber is made of
aluminum.  The polarization electrode can be set to a voltage up to
1000~V with respect to the source body.  In our setup we have somewhat
changed the original design of the polarization electrode of the
ECRIS in order to have a broad visibility of  the plasma: 
We need to optimize the X-ray source intensity to perform spectroscopy with different types of spectrometers such as Johann or double--flat crystals spectrometers.
For
that purpose, we increased the diameter of the polarization
electrode from 8 to 12~mm and made it movable, using a bellow and a
precision translation stage.  This setup allows for optimization of both  the X-ray
yield and the extracted current. Indeed, we found that even very small
changes in the position of the polarization electrode can affect
dramatically the extracted current.

 The ion transport to different experimental setups is ensured through a dedicated beam line equipped with electrostatic and magnetic optic elements. Figure \ref{fig:simpa-setup} presents a global view
of the SIMPA facility.  The solenoid is used to focus
the beam on the Faraday cup after the dipole magnet. 
The dipole allows
to select ions with a given $q/m$ ratio. In addition, varying its magnetic field, it can be used to obtained a spectrum of the charge states extracted
from the source. Its resolution is fixed by sets of slits 
located before and after it. In order to measure the
Bremsstrahlung spectra, and obtain high accuracy X-ray spectra of
the plasma, we installed either a solid-state Si(Li) detector or a
mosaic crystal spectrometer on the injection side
of the ECR ion source (see Fig.~\ref{fig:simpa-setup}).

\subsection{Set-up for the plasma densities and temperature measurements}
\label{sec:detector}
The plasma was observed through the polarization electrode, i.e., in the axial direction, at a
distance of 940~mm with a $5.755± 0.350$~mm thick crystal Si(Li)
detector, sealed with a built-in $26.5± 0.5\,\mu $m thick
beryllium window. The absolute efficiency of the detector as a
function of photon energy has been carefully measured \cite{lper2009}.

The ECRIS vacuum was separated from the atmospheric pressure by a
250~$\mu$m thick beryllium window. 
The detector was protected by a 1 cm thick lead plate, pierced by a 1~mm diameter hole to reduce the counting rate. In this geometry, the X rays emitted from the plasma are collimated by the copper polarization electrode allowing to observe a plasma region about 12~mm large. Even if relatively thin (1~mm), the copper tube results to be completely opaque for the considered photon energy \cite{has1995} because of the limited incident angle (<0.02~rad) of the X rays reaching the detector due to the finite extension of the plasma confined in a chamber of 40~mm in diameter at a distance of 940~mm from the detector (see Fig.~\ref{fig:simpa-setup}).
Pure
aluminum filters were put in front of the Si(Li)
detector in order 
to adjust and 
to keep the counting rate 
at a reasonable level
when increasing the microwave power, to avoid pile-up and worsening of
the detector response. Thicknesses of 2 to 20~mm of Al were used.

The X rays emitted by the source had also to traverse 270~mm of
air. Efficiency and absorption through filters, air and windows are
corrected for, using a Mathematica code interpolating data from
the NIST X-ray mass absorption coefficients \cite{has1995}. The
energy scale was calibrated using  an Am radioactive sample.

At this stage, it is worth mentioning that in another 
experiment \cite{kki1998},  electron temperatures
measured from an axial or a radial port of an ECR ion source have not
been found significantly different.

\subsection{Set-up for the high resolution X-ray spectroscopy}
\label{sec:mosaic-spectro}

\begin{figure}[htbp]
\centering
\includegraphics[width=\columnwidth,angle=0]{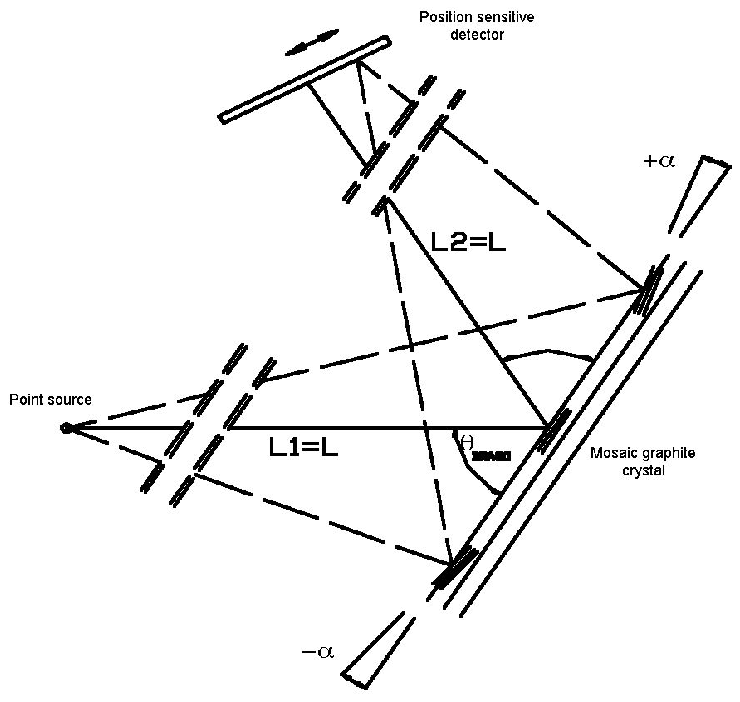}

\caption[]{Principle of the mosaic graphite flat crystal spectrometer. $\alpha$ represents the width (FWHM) of the angular distribution
of the reflecting planes in the crystal. In the present work, $L=975$~mm for Ar X rays.

}
\label{fig:spectro-mosa}
\end{figure}
To distinguish the different charge state contributions in the X-ray emission from the ions in the plasma, we used a mosaic graphite crystal spectrometer. It has been designed specifically for experiments in which good resolution as well as high efficiency are needed, such as the investigation of collision processes occurring in ion-atom \cite{rclc1985} or ion-solid \cite{srms2007} interactions, trace analysis on synchrotron radiation facilities \cite{chev1987}, and study of X-ray emission during intense laser irradiation of rare gas clusters \cite{pdlr2008}. The mosaic crystal acts as a collection of microscopic crystals, with their reflecting planes orientation following a Gaussian distribution around a mean value that corresponds to a plane parallel to the surface of the crystal. An X ray emitted by the source will always be reflected by a microscopic crystal, if its direction obeys the Bragg law corresponding to the average orientation of the crystals, within the ``mosaic width''. Our spectrometer uses a geometry that allows for a refocusing of all the X rays within the ``mosaic width'' where the source and the detector are placed symmetrically with respect to the normal of the center of the crystal and at an equal distance $L$ from the crystal (see Fig.~\ref{fig:spectro-mosa}). The working distance $L$ and the ``mosaic width'' ($0.2^\circ$) have been chosen to reach a resolution power better than 1/1000 at 3~keV.
The X rays are detected with a home-made position sensitive detector (PSD) based on a proportional counter with a resistive anode to provide the position signal. The active area of the detector is as large as $6 × 6$~cm$^2$ to entirely cover the solid angle intercepted by the crystal in its non dispersive plane. It was filled with a Xe+10\%CH$_4$ mixture at an operating pressure of 1.15~bars to ensure a detection efficiency of $\approx 75 \%$ at 3~keV \cite{pri2004}. Its spatial resolution is of 500 $\mu$m. The crystal spectrometer is under secondary vacuum ($10^{-6}$~mbar) and the detector was sealed with an aluminized (10~$\mu$g/cm$^2$) Mylar window of 12~$\mu$m thick, in order to prevent absorption of low-energy X rays. The spectrometer and associated PSD have been fully characterized by a long series of experiments, and can thus provide quantitative information on line intensities. The integrated efficiency of the spectrometer has been determined to be a few $10^{-6}$ around 3~keV (see, e.g.\cite{pri2004}) depending on the choice of $L$ distance.

The spectrometer was installed on the ECRIS axis on the injection side as shown on Fig.~\ref{fig:simpa-setup}.
The resolution of such spectrometer is directly linked to the size of the photon source in the dispersive plane of the crystal and its optimization requires a spatial definition of the plasma emission. To fulfill the focusing conditions as depicted Fig.~\ref{fig:spectro-mosa}, a 500~$\mu$m Ta diaphragm was installed inside the polarization electrode, at a position such that the diaphragm-crystal distance ($L$) was equal to the detector-crystal distance.
Use of a heavy metal, like Ta, for the diaphragm is made necessary to efficiently remove the background from the high-energy X rays coming mostly from Bremsstrahlung emitted by the plasma.
An extra 0.5~mm-wide slit was added at 105~mm from the crystal,
parallel to the dispersion plane, to reduce the counting rate. 
An X-ray filter consisting of thin metallic foils could be installed before the crystal to cut low-energy X-ray emission. A turbo-pump was installed between the ECRIS and the crystal spectrometer to ensure a good vacuum quality inside the ECRIS avoiding the use of a sealed  window.
The resolution of the spectrometer depends on the source width 
(the diameter of the Ta diaphragm in our case), on the detector resolution, on the crystal--source distance, and on the intrinsic characteristics of the crystal (i.e. the ``mosaic width'', the penetration depth of the X rays and the thickness of the crystal). In our set-up, we observed, as expected, of resolution of 2.5~eV at the Argon $K\alpha$ line energy (around 3~keV).

\section{Evaluation of the plasma characteristics}
\label{sec:plasma}
\subsection{Bremsstrahlung cross-section and emissivity formulae}
The differential Bremsstrahlung cross-section
$\sigma_{\textrm{K}}(h\nu)$ for an electron to emit a photon of energy $h\nu$ in a
collision with a nucleus of charge $Z$ is given, in a first
approximation by the Kramer's formula \cite{hut1987}:
\begin{equation}
\frac{d\sigma_{\textrm{K}}(h\nu)}{d{h\nu}}=\frac{16\pi}{3\sqrt{3}}
\alpha^3\left( \frac{\hbar}{m_e c}\right)^2
\left(\frac{c}{v_e}\right)^2 \frac{Z^2}{h\nu},
\end{equation}
where $v_e$ is the electron speed, $\alpha$ the fine structure
constant, $c$ the speed of light, $m_e$ the mass of the electron.

In an ECR plasma, Bremsstrahlung cross-sections are measured from the emissivity density $J(h\nu)= h\nu\ N(h\nu)$
where $N(h\nu)$ is the plasma spectral function that represents
the total number of photons of energy $h\nu$ emitted by the plasma
per unit of time, volume and energy. Both $J$ and $N$ are related to the
Bremsstrahlung cross-section and to the 
ionic and electronic densities $N_i$ and $N_e$ by the formula
\begin{equation}
J(h\nu) =  N_{i}N_{e} h\nu \int_{h\nu}^{\infty} \frac{d\sigma_{\textrm{K}}(h\nu)}{d{h\nu}} v_e(E) f(E) dE.
\label{eq:redemissivity}
\end{equation}
Here $f(E)$ is the electron energy distribution, $v_e(E)$ the speed of an electron of energy $E$. 

Experimentally, $J(h\nu)$ is determined from measured spectra $N^\text{ch.}(h\nu)$ corresponding to the number of counts per channel accumulated in an integration time $t$ in a set-up with a solid angle $\Omega$,
\begin{equation}
J(h\nu)= h\nu\ \frac{N^\text{ch.}(h\nu)}{\eta(h \nu)\ t} \frac{4 \pi} {\Delta E\ V\ \Omega} ,
\label{eq:redemissivity_exp}
\end{equation}
where $\eta(h \nu)$ is the efficiency of the experimental apparatus for an X ray of energy $h \nu$, $\Delta E$ is the energy interval corresponding to a channel and $V$ is the portion of the plasma volume seen by the detector.

The electrons inside an ECRIS plasma have been shown to obey a
non-Maxwellian distribution (see, e.g. \cite{dkgb2000}). This distribution can be represented
as the sum of two Maxwellian distributions $f_{\textrm{Mw}}(E)$ corresponding
to a low-energy and high-energy electron populations, where
\begin{equation}
f_{\textrm{Mw}}(E)=\frac{2}{\sqrt{\pi}}
\frac{E^{1/2}}{\left(kT_{\textrm{Mw}}\right)^{3/2}}
e^{\left(-\frac{E}{kT_{\textrm{Mw}}}\right)}.
\label{eq:maxwell}
\end{equation}
$T_{\textrm{Mw}}$ is the Maxwellian temperature and $k$ the
Boltzmann constant.
The high-energy component has a typical temperature of several tens of keVs
\cite{blbg1994} while the low-energy component has a much lower value, around 1~keV. 
Measured temperatures for the high energy
electron population range from 5 to 13~keV for an ECRIS
operated at 2.45~GHz \cite{ghbb1998,kki1998}, 40--60~keV for a CAPRICE
source operated at 10~GHZ \cite{dkgb2000}, and up to 100~keV for
an ECRIS operated at 18 and 28~GHz \cite{blbg1994,lblt2008}.
As in the measurements cited above, our study is sensitive to the high-energy electron component
due to our detection energy range of $\sim 20-120$ ~keV.
In the following paragraphs, we will refer only to such a component.

Equation (\ref{eq:maxwell}) does not take into account
relativistic effects in the electron motion, and its use for the
high-energy component of the electron population, where $k T \lesssim m_e c^2$  is approximate.
A more realistic distribution that takes into account relativistic effects is the Maxwell--Jüttner distribution \cite{syn1957},
\begin{equation}
  f_{\textrm{MJ}}(\gamma)= \frac{m_ec^2}{kT_{MJ}} \frac{\gamma^2 \beta}{K_2\left(\frac{m_ec^2}{kT_\textrm{MJ}}\right)}
e^{\left(-\frac{m_ec^2 \gamma }{kT_{\textrm{MJ}}}\right)},
\label{eq:maxwell-juttner}
\end{equation}
where $\beta$ and $\gamma=1/ \sqrt{1-\beta^2}$ are the relativistic factors related to the kinetic energy $E$ and the velocity $v_e$ by the relationships
$\gamma(E) = 1 + \frac E {m_ec^2}$ and $v_e = \beta c$.
$K_2(x)$ is the modified Bessel function of the second kind.
With the relativistic notation, Eq.~(\ref{eq:redemissivity}) becomes:
\begin{equation}
J(h\nu) = N_{i}N_{e} h\nu \int_{\gamma(h\nu)}^{\infty} \frac{d\sigma_{\textrm{K}}(h\nu)}{d{h\nu}} v_e(\gamma) f_{\textrm{MJ}}(\gamma) d\gamma.
\label{eq:redemissivity_rel}
\end{equation}

In the case of the
Maxwellian  distribution without relativistic effects, Eq.~(\ref{eq:maxwell}), and using Kramer's cross-section, the integral
in Eq.(\ref{eq:redemissivity}) is elementary and gives
\cite{laa1997}:
\begin{multline}
J_{\textrm{M-K}}(h\nu) = N_{i}N_{e} (Z\hbar)^2\left(\frac{4\alpha}{\sqrt{6
m_e}}\right)^3  \\
  × \left( \frac{\pi}{kT_{\textrm{Mw}}}\right)^{1/2}
e^{\left(-\frac{h\nu}{kT_{\textrm{Mw}}}\right)}.
\label{eq:emissivity_int}
\end{multline}

Taking into account the relativistic formula, the integral in Eq.~(\ref{eq:redemissivity_rel})
gives:
\begin{multline}
J_{\textrm{MJ-K}}(h\nu)= N_{i}N_{e} 16 \pi \frac{(Z\hbar)^2}{K_2\left(\frac{m_ec^2}{kT_\textrm{MJ}}\right)}
\left(\frac{\alpha\ c}{\sqrt{3}}\right)^3 \left( \frac 1 {m_e c^2} \right)^4  \\
 × \Big[ (m_ec^2)^2+2(kT_{\textrm{MJ}})^2+2 kT_{\textrm{MJ}} h\nu + \\
(h\nu)^2 + 2 m_ec^2 kT_{\textrm{MJ}} + 2 m_ec^2 h\nu \Big]
e^{\left(-\frac{m_ec^2+h\nu}{kT_{\textrm{MJ}}}\right)},
\label{eq:emissivity_int_rel}
\end{multline}
which can be written as power expansion in $kT/(mc^2)$:
\begin{multline}
J_{\textrm{MJ-K}}(h\nu)= 
N_{i}N_{e} (Z\hbar)^2 \left(\frac{4\alpha}{\sqrt{6
m_e}}\right)^3  \\
  × \left( \frac{\pi}{kT_{\textrm{Mw}}}\right)^{1/2}
e^{\left(-\frac{h\nu}{kT_{\textrm{Mw}}}\right)}
× \Big[ 1 + \frac{16 h\nu + kT_{\textrm{MJ}}}{8 m_ec^2} + \\
\frac{128 (h\nu)^2 -224\ h\nu\  kT_{\textrm{MJ}} +121 ( kT_{\textrm{MJ}})^2}{128 (m_ec^2)^2} \Big] \\
+ \mathcal{O}\left[ \left(\frac {kT} {m_ec^2} \right)^2 \right].
\label{eq:emissivity_int_dev}
\end{multline}
As it can be noted, the zeroth order corresponds well to Eq.~(\ref{eq:emissivity_int}), which is a valid approximation for $kT/(m_ec^2)\ll 1$.

In the following section, Eq.~\eqref{eq:emissivity_int_dev} will be used to extract the $N_eN_i$ product and the electronic temperature of the plasma as well, which refers to a ``spectral temperature'' as defined in Ref. \cite{lltv2006}. However, it is worth mentioning that applying a relativistic Maxwell distribution to fit the observed emissivity in the high energy range, instead of what has been done so far in the literature where non-relativistic formula have been used, should lead to temperature values closer to the ``real average temperature'' of the electrons in the plasma since those electrons are energetic. The temperature value extracted from the fit using Eq.~\eqref{eq:emissivity_int_dev} is 15\% smaller than in a non-relativistic analysis.

\subsection{Measurement of the electronic  spectral temperature}
\label{sec:temp}
\begin{figure}[htbp]
\centering
\includegraphics[width=\columnwidth,angle=0]{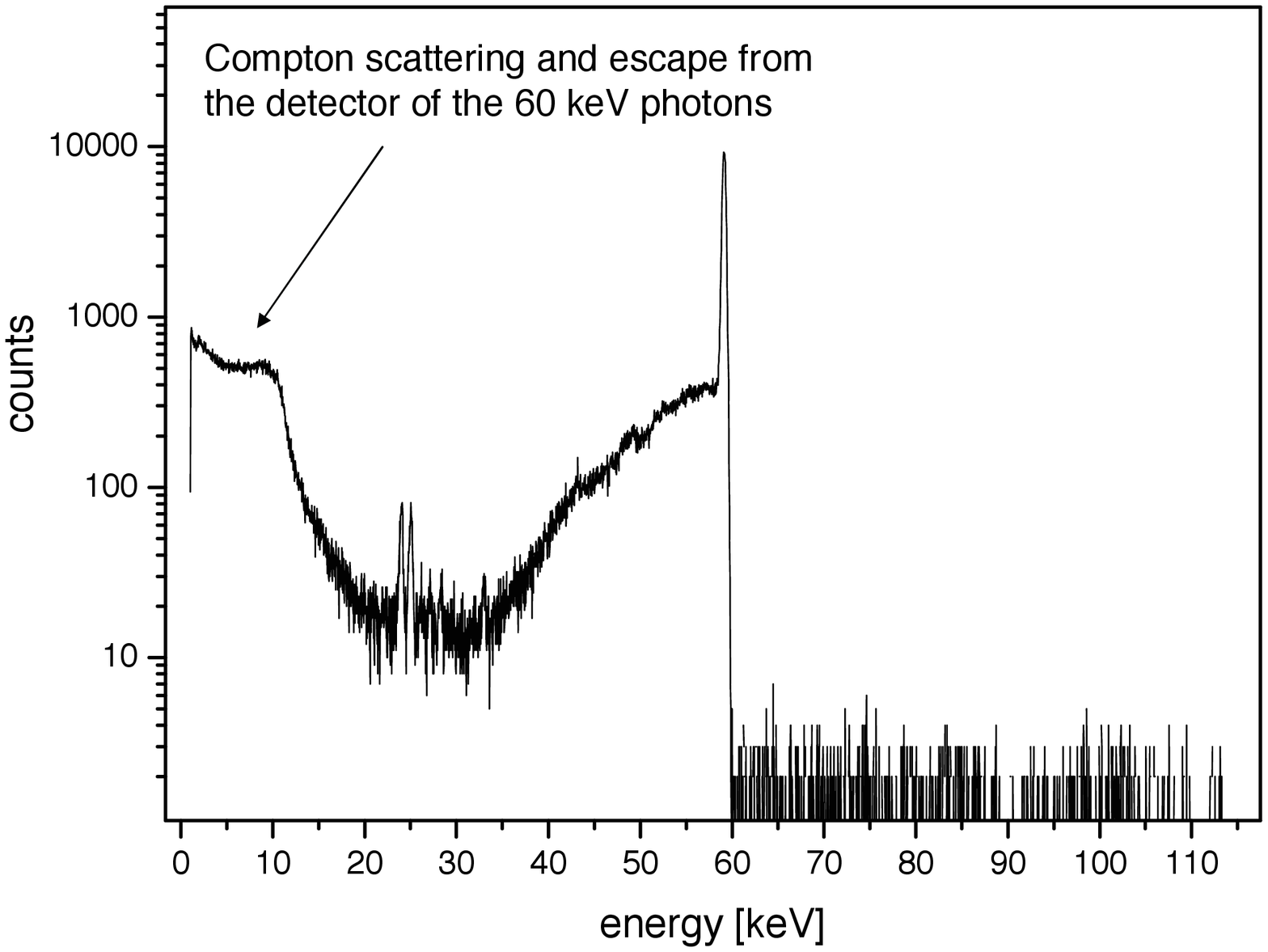}

\caption[]{Spectrum recorded by Si(Li) detector with the Am calibration source with a
thick aluminum filter (20 mm) in front.The 60~keV americium $\gamma$-line
and the associated Compton scattering signal
are clearly recognizable.}
\label{fig:am-spectr}
\end{figure}

Different gases were injected into the ion source, argon, krypton and xenon, as well as oxygen rare-gas mixture. Other significant gases present in the source were nitrogen and water. 
The injected pressure was always in the order of $p=10^{-5}$--$10^{-4}$~mbar. Note, that $p$ is the neutral gas pressure as measured at the  gas injection position and not the pressure inside the plasma chamber.
 During the measurement,  only the power of 
 the injected  microwave was changed. 
A typical acquisition time of 1000-3000~sec was required to obtain valuable Bremsstrahlung spectra with the solid state detector.
\begin{figure}[htbp]
\centering
\includegraphics[width=\columnwidth,angle=0]{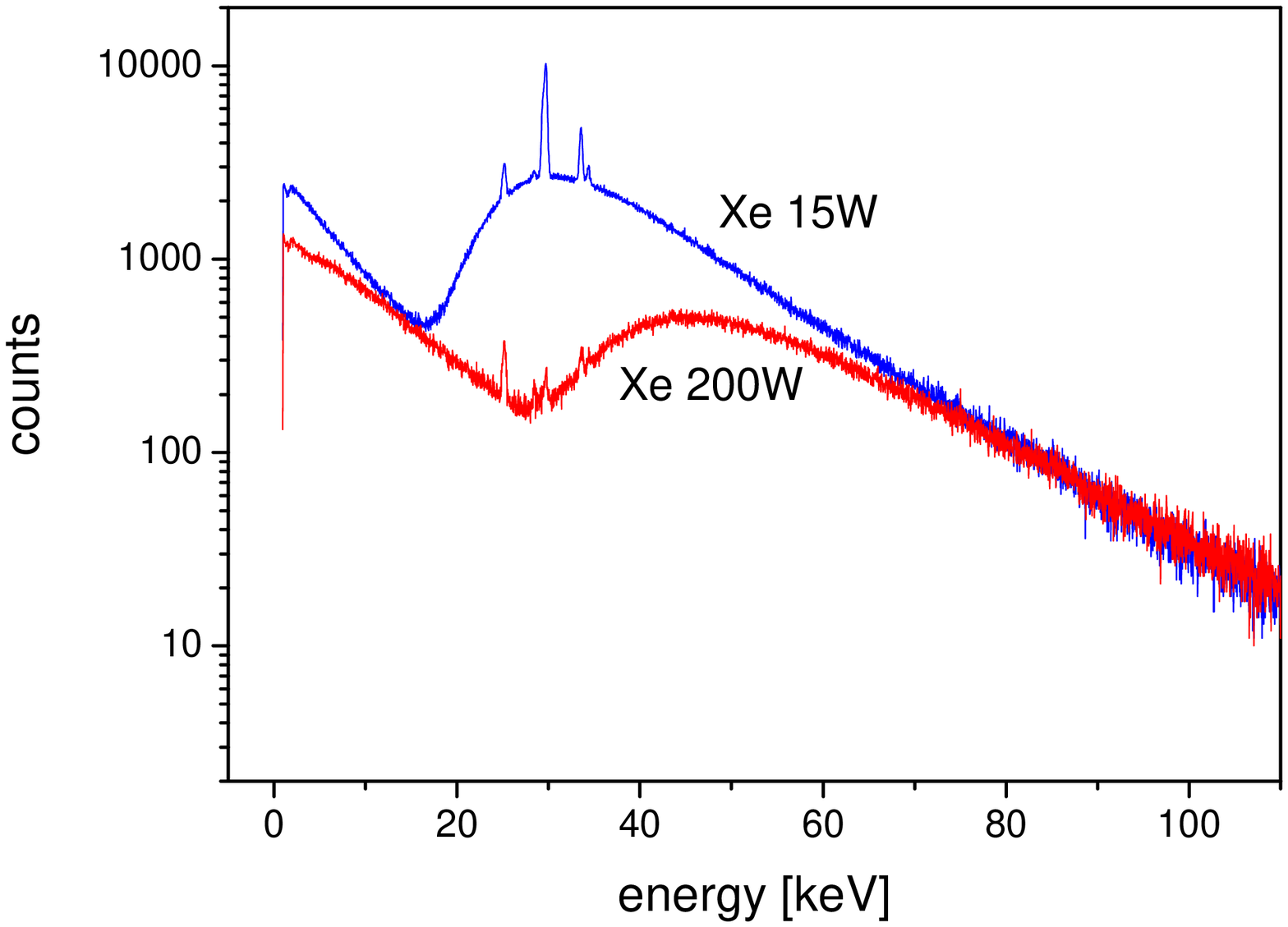}

\caption[]{ (Color online) Xe X-ray spectra obtained with two different microwave
powers applied to the source. Energy scale is obtained using an Am
source. The Xe $K\alpha$, $K\beta$ and $K\gamma$ lines can be
observed. Filter thickness used were 4~mm at 15~W  and 20~mm at
200~W.} \label{fig:raw}
\end{figure}
We present on Fig.~\ref{fig:raw}  typical raw spectra obtained with the solid state detector for a Xe plasma at two different injected microwave powers. The Bremsstrahlung tail clearly appears up to energies of about 110~keV. 
Even with filters that should cut completely the low energy part of the spectrum, one observe an increase of the intensity at low energy. This increase is, in fact, due to Compton scattering and escape of high energy photons from the Si crystal,
 which leads to only a partial deposition of the photon energy in the detector. Therefore, such events produce counts in the low energy part of the spectra. This can be seen by examining spectra for the Am calibration source
(Fig.~\ref{fig:am-spectr}) recorded with a thick Al absorber in front of the detector so that only the 60 keV photons could pass through.

\begin{figure}[htbp]
\centering
\includegraphics[width=\columnwidth,angle=0]{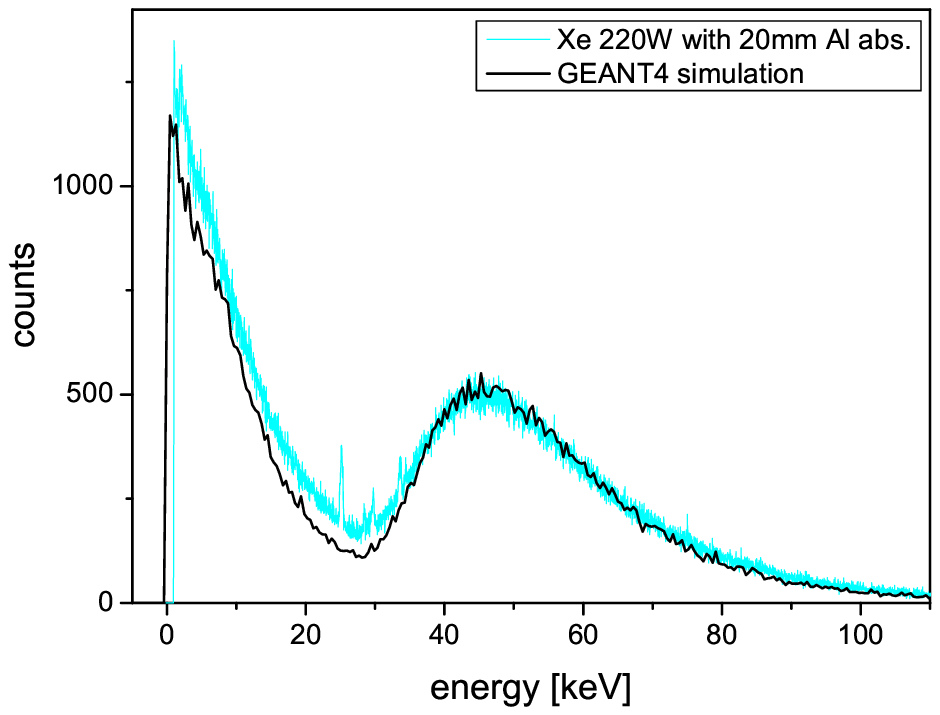}
\caption[]{(Color online) A comparison between the spectra for Xe as observed in
the experiment (220~W: 20~mm Al) and the
corresponding GEANT~4 simulation.} \label{fig:geant4-sim}
\end{figure}

In addition to this experimental test, a simulation with 
GEANT~4.8.3  \cite{geant4} has been performed to study the 
effect of the Compton scattering.
In the simulation, the identical experimental conditions were considered, i.e., crystal and window sizes, material, thicknesses and specifics of the setup geometry.

As can be observed on Fig.~\ref{fig:geant4-sim}, the simulation
clearly reproduces the low-energy intensity divergence observed in
experimental spectra (as in Fig.~\ref{fig:raw}) which is intrinsic to using silicon as crystal for detection.

\begin{figure}[htbp]
\centering
\includegraphics[width=\columnwidth,angle=0]{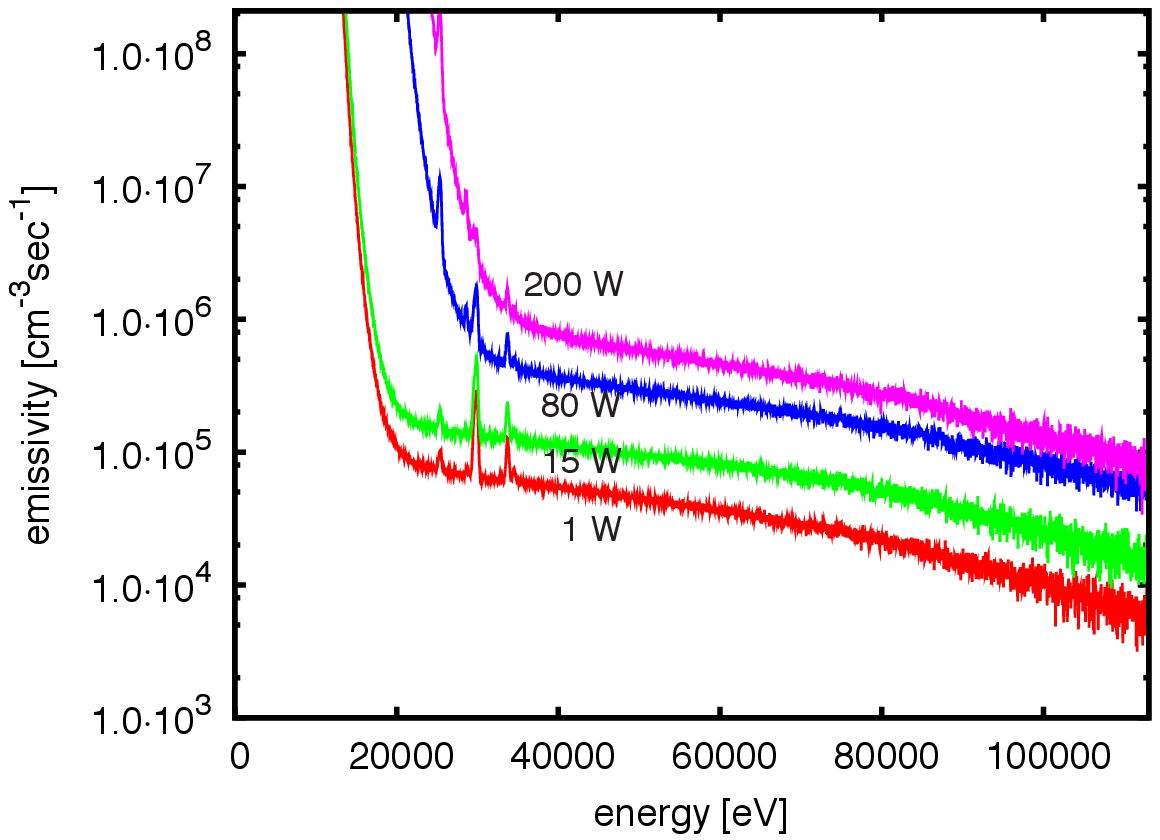}
\caption[]{(Color online) Examples of emissivity spectra obtained with different
microwave powers applied to the source with Xe. Each spectrum is corrected
for detector efficiency, air, filter and windows absorption. All
spectra have been normalized to the acquisition time, corrected
for dead-time. The Xe
$K\alpha$, $K\beta$ and $K\gamma$ lines can be observed, even
through thick filters. Filter thickness are: at 1~W and 15~W: 4~mm, at 80~W and at 200~W: 20~mm.}
\label{fig:cor-bress}
\end{figure}
The corresponding emissivity spectra obtained for a Xe plasma at different injected microwave powers are shown on Fig.~\ref{fig:cor-bress}, after correction for detector efficiency and filter absorption according to Eq.~(\ref{eq:redemissivity_exp}).
To evaluate the temperature, we perform a fit of the experimental emissivity in a range of 40~keV to 70~keV, using Eq.~(\ref{eq:emissivity_int_dev}) with the density product $N_i N_e$ and the temperature $T_\textrm{MJ}$ as free parameters.

The lower limit of the energy range has been chosen to avoid any influence of the Compton effect. 
The upper limit is given by the photon energy for which, both, 
the detector efficiency becomes too small and the statistical
uncertainty too large. 
The weights in the emissivity fit are evaluated by taking, at each point of the spectrum, the square root of the measured intensity $N^\text{ch.}(h\nu)$. The non-linear fit has been performed with two different codes
 to cross-check the results. We developed an in-house code, using standard techniques \cite{pftv1986}, and we also used the MINUIT CERN library \cite{jaro1975}.

\begin{figure}[htbp]
\centering
\includegraphics[width=\columnwidth,angle=0]{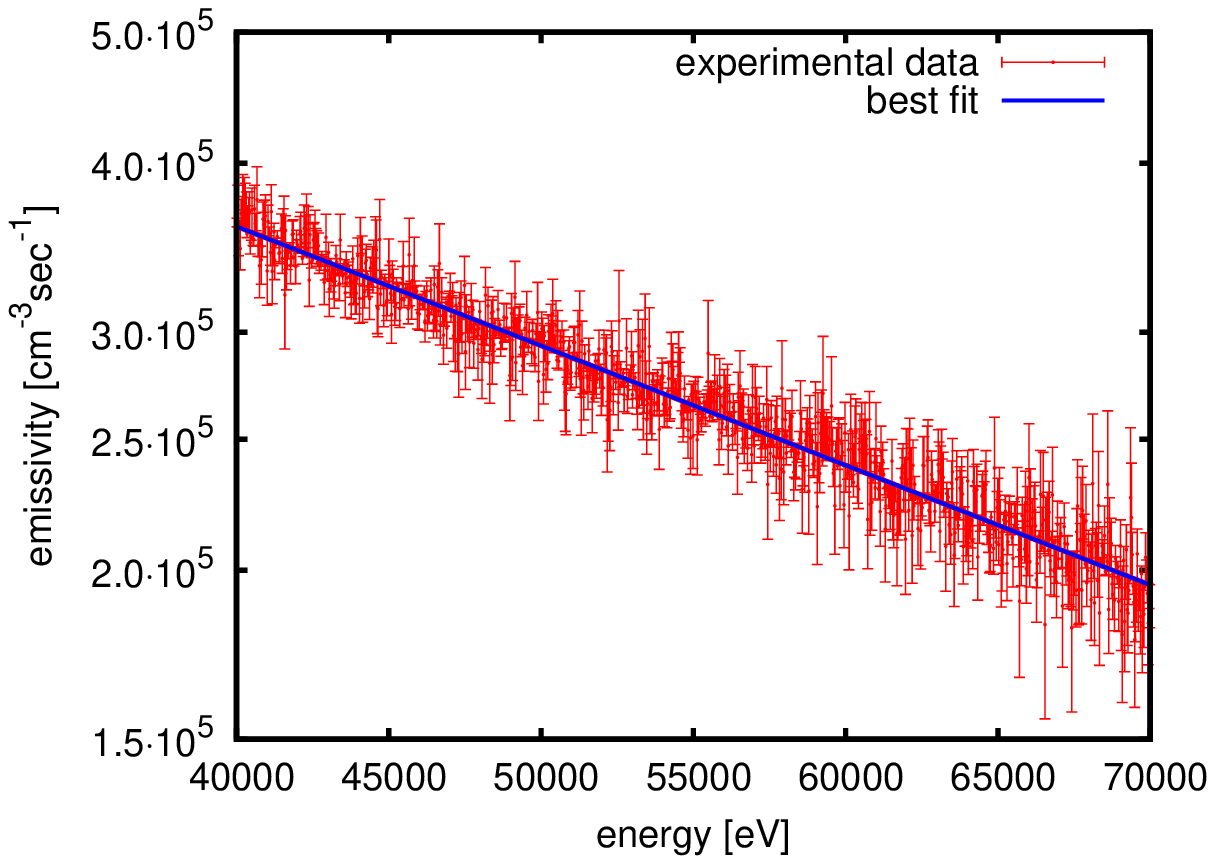}

\caption[]{(Color online) Emissivity spectrum by a Xe plasma at a microwave
power of 80~W in logarithmic  scale along
with the fit.

} \label{fig:xe-fit-spect}
\end{figure}

An example of fit obtained for Xe plasma with an injected power of 80~W is presented on Fig.~\ref{fig:xe-fit-spect}. The injection pressure was $5.1× 10^{-5}$~mbar and the acquisition time
2848~s (dead-time corrected).
The  spectral temperature and density product
are found to be $T_\textrm{MJ}=42.27 ± 0.21$~keV and $N_i N_e = (5.82 ± 0.02) × 10^{23}$~part$^2$/cm$^6$, taking into account here, only the statistical errors.
The corresponding reduced $\chi^2$ is equal to 0.3 for a fitting range between 40~keV and 70~keV.

Indeed, when same plasma conditions are applied, temperature values extracted from measurements with different absorber thicknesses deviate from each other by up to 10\%.
Concerning absorption corrections,
it is worth mentioning that corrections based on \cite{has1995} did neither
completely nor consistently compensate for the absorbers of  different
thicknesses.  
This is due, on the one hand, to uncertainty on the filter thickness, 
and on the other hand to the 
fact that we have an extended photon source with a diameter of 12~mm  (corresponding to  the polarization electrode aperture) and not a ``narrow'' and well collimated photon beam. 
Consequently, a small fraction of photons scattered in the
absorber at certain angles can  pass through the Pb
collimator hole
and hit the detector, producing  additional counts 
not taken into account by the
absorber corrections. 
Another source of uncertainty is due to the absorption correction, as given by
\cite{has1995}: a deviation between experimental values and
theoretical tables have been exhibited by recent direct measurements on Si
\cite{tcb2003,tcbp2003} and other elements \cite{jtcb2005,jtcb2007}.
The measurements for Si, however, cover a much lower energy range
, and cannot be used to improve our value. In that respect, we performed specific measurements to obtain the overall corrections due
to all these absorption effects. More precisely, to account for, we extract correction factors 
comparing data obtained for different Al
absorber thicknesses and without any absorber at a fixed power of 
1~W (the photon rate being  too high without absorber at higher power).
A linear interpolation of those data allows us to apply correction factors that lead to reliable values of plasma temperatures and densities. 
The uncertainty due to the linear interpolation is taken into account in the calculation of the total uncertainty.

The limited accuracy on the detector crystal thickness, equal to $5755 ± 350~\mu$m \cite{lper2009}, introduces an additional source of systematic uncertainty on detector efficiency, which corresponds to about 5\% and 
8\% of the $T_\textrm{MJ}$ and $N_i N_e$ values, respectively.

\begin{figure}[htbp]
\centering
\includegraphics[width=\columnwidth,angle=0]{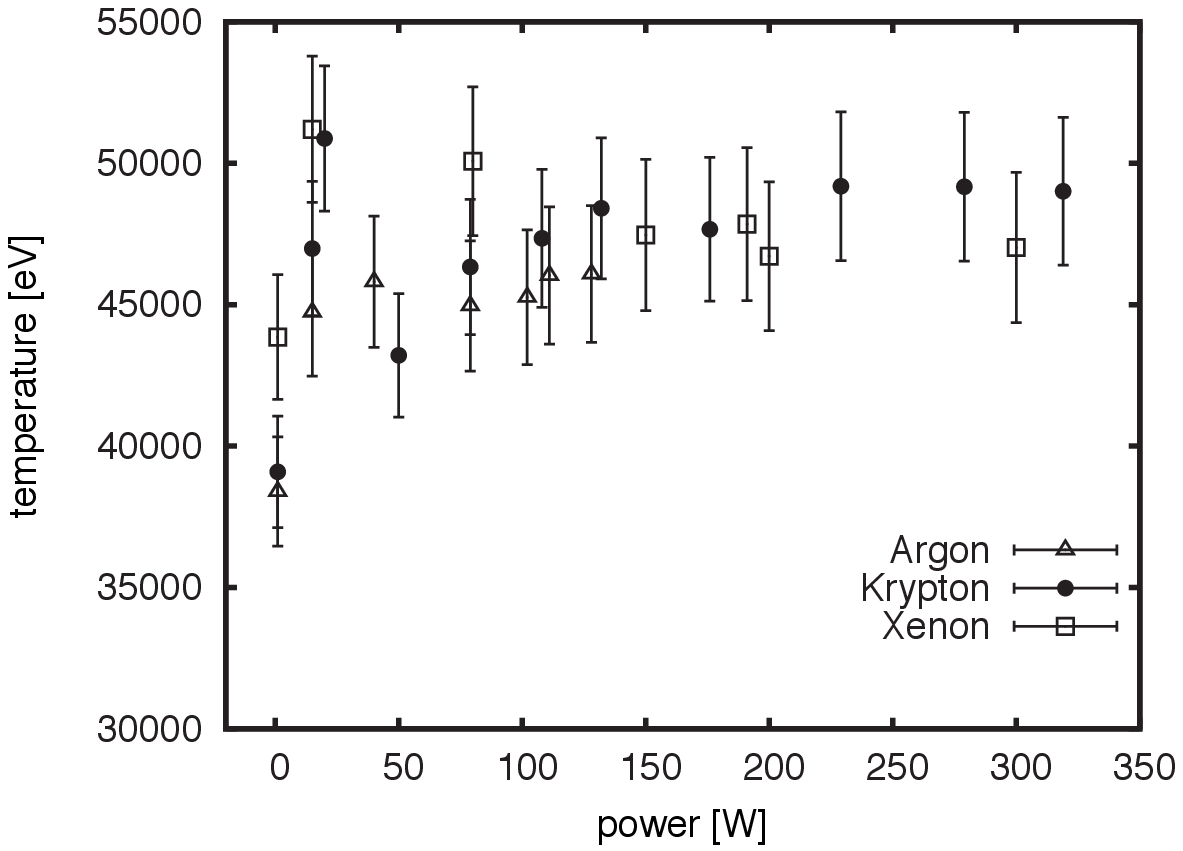}
\caption[]{ Electronic spectral temperature
$kT_{\textrm{MJ}}$  in eV, as a function of the microwave power in SIMPA and for different injected gases.

} \label{fig:temp-simpa}
\end{figure}

In Tables~\ref{tab:ar},\ref{tab:kr} and \ref{tab:xe} and Fig.~\ref{fig:temp-simpa}, we present the extracted temperatures as a function of microwave power for different gases (argon, krypton, xenon). 
The temperature for Ar and Kr is around 39~keV at low power and a stabilized temperature of 46--49~keV is reached at powers around 100~W (Fig.~\ref{fig:temp-simpa}). For Xe however, the temperature is more or less constant around 48~keV.

In contrast to Refs.~\cite{kki1998,blbg1994}, we observed no particular dependence on the injected pressure value in the range $p=1-10× 10^{-5}$~mbar.
We did not observe either any dependence on the injected gas composition.
Even the use of an oxygen--argon mixture instead of argon alone did not change the measured electron temperature.

\begin{table}
\caption{Electronic spectral temperatures, ionic and electronic density products of the plasma for argon gas as a function of the injected microwave power.
All values include systematic corrections and uncertainties described in the text.}
\label{tab:ar}
\begin{tabular}{c c c}
\hline
power [W] & $T_\textrm{MJ}$ [KeV] & $N_{i}N_{e}$ [$10^{22}$ part$^2$/cm$^6$]\\
\hline
1	& $	38.4	±	3.1	$ & $	7.0	±	0.6	$ \\
15	& $	44.7	±	3.6	$ & $	16.2	±	1.3	$ \\
40	& $	45.8	±	3.7	$ & $	30.4	±	2.6	$ \\
79	& $	45.0	±	3.6	$ & $	57.7	±	5.2	$ \\
102	& $	45.3	±	3.7	$ & $	79.1	±	7.6	$ \\
111	& $	46.0	±	3.8	$ & $	42.2	±	4.1	$ \\
128	& $	46.1	±	3.8	$ & $	97.2	±	9.4	$ \\
\hline
\end{tabular}
\end{table}

\begin{table*}
\caption{Electronic spectral temperatures, ionic and electronic densities and ion mean charge state  of the plasma for krypton gas as a function of the injected microwave power.
All values include systematic corrections and uncertainties described in the text.}
\label{tab:kr}
\begin{tabular}{c c c c c c}
\hline
power [W] & $T_\textrm{MJ}$ [KeV] & $N_{i}N_{e}$[$10^{22}$ part$^2$/cm$^6$] &  $<q>$ & $N_{i}$[$10^{11}$ part/cm$^3$] & $N_{e}$[$10^{11}$ part/cm$^3$] \\
\hline
1	& $	39.1	±	2.0	$ & $	5.2	±	0.4	$ & $	4.9	±	0.2	$ & $	1.03	±	0.05	$ & $	5.04	±	0.24	$ \\
15	& $	47.0	±	2.4	$ & $	12.8	±	1.0	$ & $	6.8	±	0.2	$ & $	1.37	±	0.06	$ & $	9.35	±	0.41	$ \\
20	& $	50.9	±	2.6	$ & $	16.4	±	1.3	$ & $				$ & $								$ \\
50	& $	43.2	±	2.2	$ & $	90.7	±	7.4	$ & $	7.9	±	0.1	$ & $	3.40	±	0.14	$ & $	26.69	±	1.12	$ \\
79	& $	46.3	±	2.4	$ & $	59.6	±	5.5	$ & $	8.2	±	0.2	$ & $	2.70	±	0.13	$ & $	22.08	±	1.04	$ \\
108	& $	47.3	±	2.4	$ & $	79.1	±	7.3	$ & $	8.3	±	0.2	$ & $	3.09	±	0.15	$ & $	25.60	±	1.21	$ \\
132	& $	48.4	±	2.5	$ & $	102.0	±	9.5	$ & $	8.4	±	0.2	$ & $	3.48	±	0.17	$ & $	29.31	±	1.41	$ \\
176	& $	47.7	±	2.5	$ & $	115.6	±	11.5	$ & $	8.2	±	0.4	$ & $	3.75	±	0.20	$ & $	30.85	±	1.67	$ \\
229	& $	49.2	±	2.6	$ & $	125.9	±	12.5	$ & $				$ & $								$ \\
279	& $	49.2	±	2.6	$ & $	132.6	±	13.2	$ & $				$ & $								$ \\
319	& $	49.0	±	2.6	$ & $	163.4	±	16.2	$ & $				$ & $								$ \\
\hline
\end{tabular}
\end{table*}

\begin{table*}
\caption{Electronic spectral temperatures, ionic and electronic densities, and ion mean charge state  of the plasma for xenon gas as a function of the injected microwave power.
All values include systematic corrections and uncertainties described in the text.}
\label{tab:xe}
\begin{tabular}{c c c c c c}
\hline
power [W] & $T_\textrm{MJ}$ [KeV] & $N_{i}N_{e}$[$10^{22}$part$^2$/cm$^6$] & $<q>$ & $N_{i}$[$10^{11}$ part/cm$^3$] & $N_{e}$[$10^{11}$ part/cm$^3$] \\
\hline
1	& $	43.9	±	3.5	$ & $	9.3	±	0.8	$ & $	7.8	±	0.5	$ & $	1.09	±	0.06	$ & $	8.56	±	0.44	$ \\
15	& $	51.2	±	4.1	$ & $	18.2	±	2.3	$ & $	8.2	±	0.4	$ & $	1.49	±	0.10	$ & $	12.24	±	0.83	$ \\
80	& $	50.1	±	4.1	$ & $	69.4	±	13.3	$ & $	9.6	±	0.2	$ & $	2.69	±	0.26	$ & $	25.79	±	2.49	$ \\
150	& $	47.5	±	4.0	$ & $	134.2	±	23.2	$ & $	9.3	±	1.0	$ & $	3.79	±	0.39	$ & $	35.41	±	3.60	$ \\
191	& $	47.8	±	4.0	$ & $	140.2	±	22.5	$ & $				$ & $								$ \\
200	& $	46.7	±	3.9	$ & $	166.3	±	26.3	$ & $				$ & $								$ \\
300	& $	47.0	±	4.0	$ & $	215.8	±	29.3	$ & $				$ & $								$ \\
\hline
\end{tabular}
\end{table*}

\subsection{Ionic and electronic densities and mean charge state of ions }
\label{sec:dens}
\begin{figure}[htbp]
\centering
\includegraphics[width=\columnwidth,angle=0]{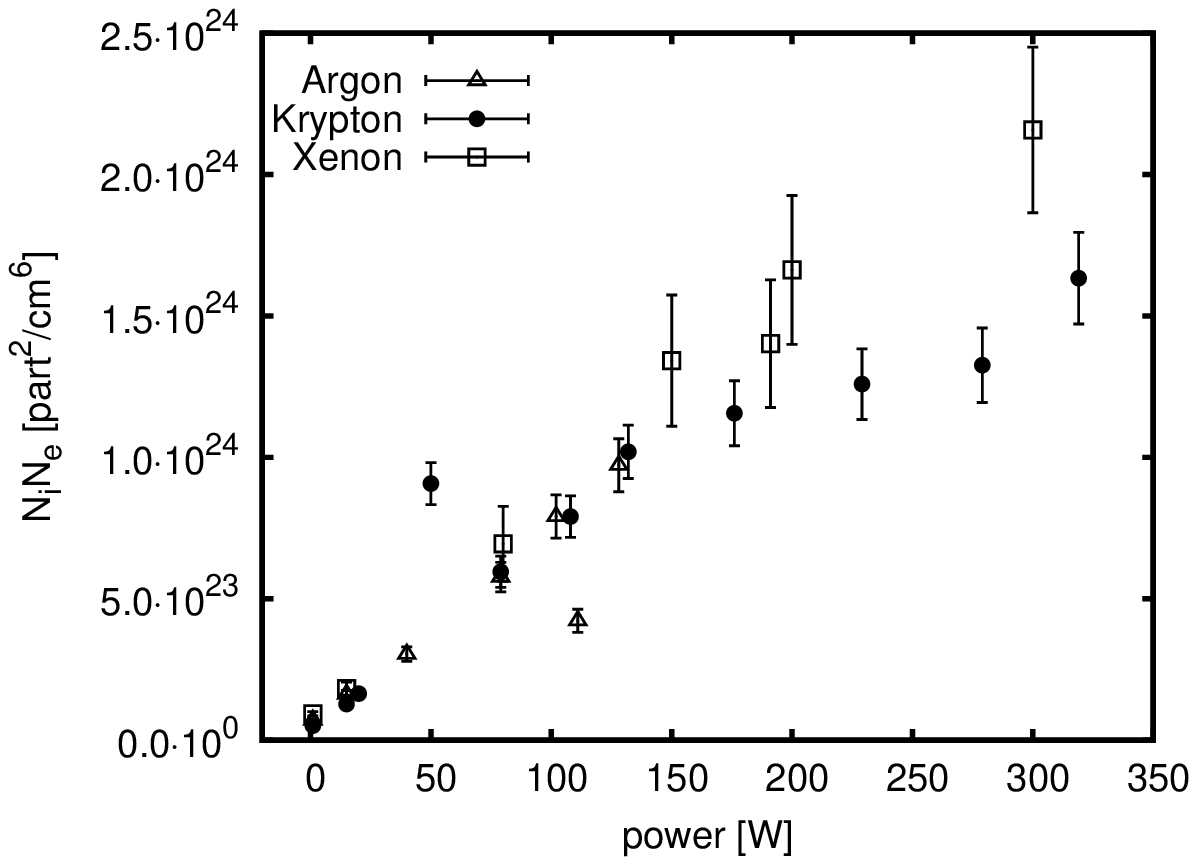}

\caption[]{Ionic and electronic density product as a function of injected power and for different gases.}
\label{fig:NiNe}
\end{figure}

Additionally to the electron temperature, another important parameter could be extracted from the Bremsstrahlung spectra measurement, namely the product of the electronic and ionic densities $N_i N_e$.  
From Eqs.~(\ref{eq:emissivity_int}) and (\ref{eq:emissivity_int_rel}), we see that the emissivity depends linearly on the  $N_{i}N_{e}$ product and its value was deduced from the emittance fit.
In our set-up the detected Bremsstrahlung radiation was produced principally from the plasma. 
It is worth mentioning that contribution of emission by the electrons hitting the chamber wall is also possible. Nevertheless, any significant effect is quite unlikely with  our set-up since,  
as can be observed on Fig.~\ref{fig:raw}, no fluorescence emission from the stainless steel wall is present (no iron fluorescence lines).
The values of the measured density product are presented in Tables~\ref{tab:ar},\ref{tab:kr} and \ref{tab:xe} 
and in Fig.~\ref{fig:NiNe}.
In contrast with the electronic temperature, the $N_iN_e$ value in the source changes more or less linearly with the microwave power
(Fig.~\ref{fig:NiNe}), whilst  no particular dependency on the gas composition has been noticed, at least below 150W.

\hfill

Simultaneously to some of the X-ray spectra acquisition with Kr and Xe gases, a measurement of the extracted ionic current was performed. This detection allows for the determination of the mean ionic charge state $ \langle q \rangle$ of the plasma, which is also a good indication of the source performance.
In addition,  simultaneous measurements of $ \langle q \rangle$ and $N_i N_e$ allow to disentangle the  $N_i N_e$ product itself, and lead to separate estimations of the electronic and ionic densities. Indeed, since we are observing a neutral plasma, we have $N_e \simeq \langle q \rangle \  N_i$, which implies that  $N_i \simeq   \sqrt{ N_iN_e /\langle q \rangle }$ and $N_e \simeq \sqrt{\langle q \rangle\ N_iN_e}$.

To determine  the mean ion charge of the plasma, we make the likely hypothesis that the extracted current $I_q^{(i)}$ of a given charge state $q$ and species $i$ is proportional to the number of ions in that charge state in the plasma.
It is shown that (see, \cite{hmg2000} and references therein)
\begin{equation}
I_q^{(i)} \propto \frac{1}{2} \frac{n_q^{(i)} q e V_p}{\tau_q^{(i)}},
\label{eq:ionic-curr}
\end{equation}
where $n_q^{(i)}$ is the density of ions of charge $q$ for a given species
$i$, $\tau_q^{(i)}$ the ion confinement time, and $V_p$ the volume of
the hot plasma under the influence of the extraction electrode
electric field. The hot plasma volume is roughly equal to the volume
contained in the surface of a constant $B$-field in a given 
ECR ion source.  
The measurement of the different $I_q^{(i)} $ was obtained using the dipole coupled to a Faraday cup (see Fig.~\ref{fig:simpa-setup}).
The range of the measurable charge states was limited for the higher charge states by the minimum detectable ionic current, $I_q^{(i)} \gtrsim 0.1~\mu$A.
The lowest accessible charge state was determined by the maximum magnetic field of the dipole and the extraction voltage.

To deduce $\langle q \rangle$ from the different $I_q^{(i)} $, we apply an additional approximation in Eq.~\eqref{eq:ionic-curr}. 
As it has been measured in Ref.~\cite{dkgb2000}, the different values of $\tau_q^{(i)}$, are slightly dependent on the charge state of the ion.
For our propose, we assume that the ion confinement time does not depend on $q$ and $I_q^{(i)} \propto n_q^{(i)} q $.
With this approximation, the sum of counts in each charge state peak was calculated and the charge state spectra were extrapolated to include lower and higher charge states that were not measured by assuming a Gaussian distribution. For Kr the charge state intensities from 6+
to 15+ were recorded, whereas it was limited from  7+ to 15+ in the case of Xe.

\begin{figure}[htbp]
\centering
\includegraphics[width=\columnwidth,angle=0]{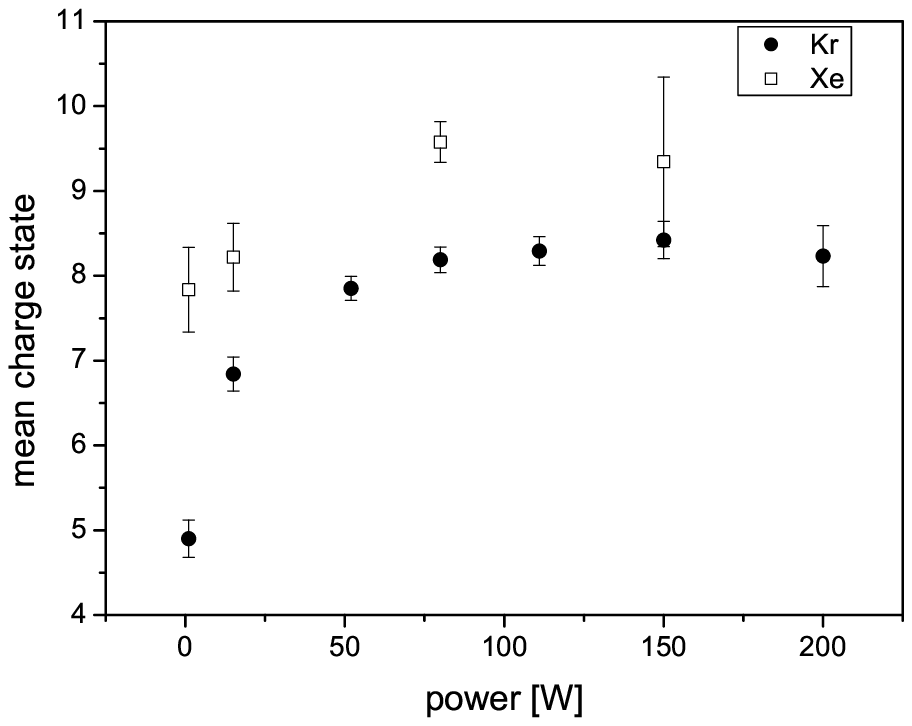}

\caption[]{Mean charge state as a function of injected power for Kr and Xe ions.}
\label{fig:mean-charge}
\end{figure}

 Results on mean charge states are presented on Fig.~\ref{fig:mean-charge} and Tables~\ref{tab:kr} and \ref{tab:xe}.
A value of 8+ for Kr ions and 9+ for Xe ions has been observed, with a weak dependence of the injected power above 50W.

\begin{figure}[htbp]
\centering
\includegraphics[width=\columnwidth,angle=0]{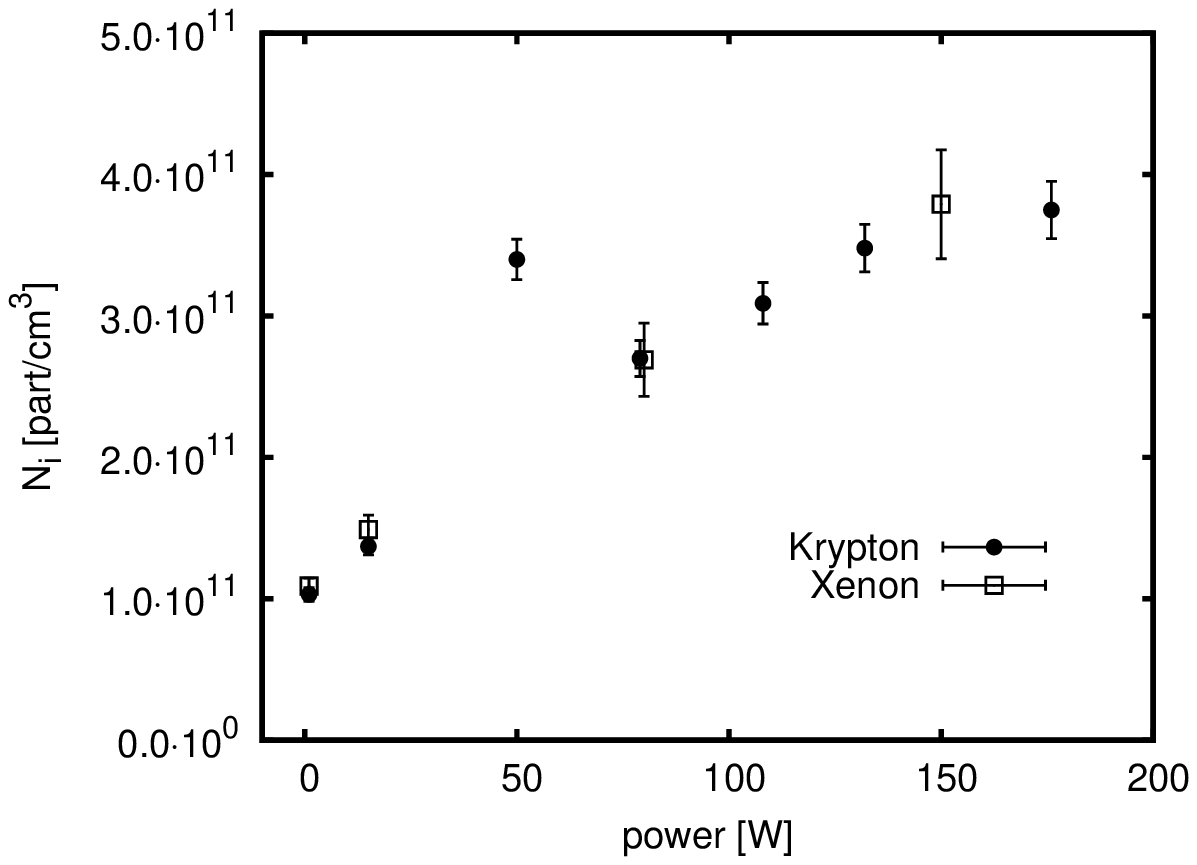}

\caption[]{Ionic density for Kr and Xe gases as a function of injected power.}
\label{fig:Ni}
\end{figure}

\begin{figure}[htbp]
\centering
\includegraphics[width=\columnwidth,angle=0]{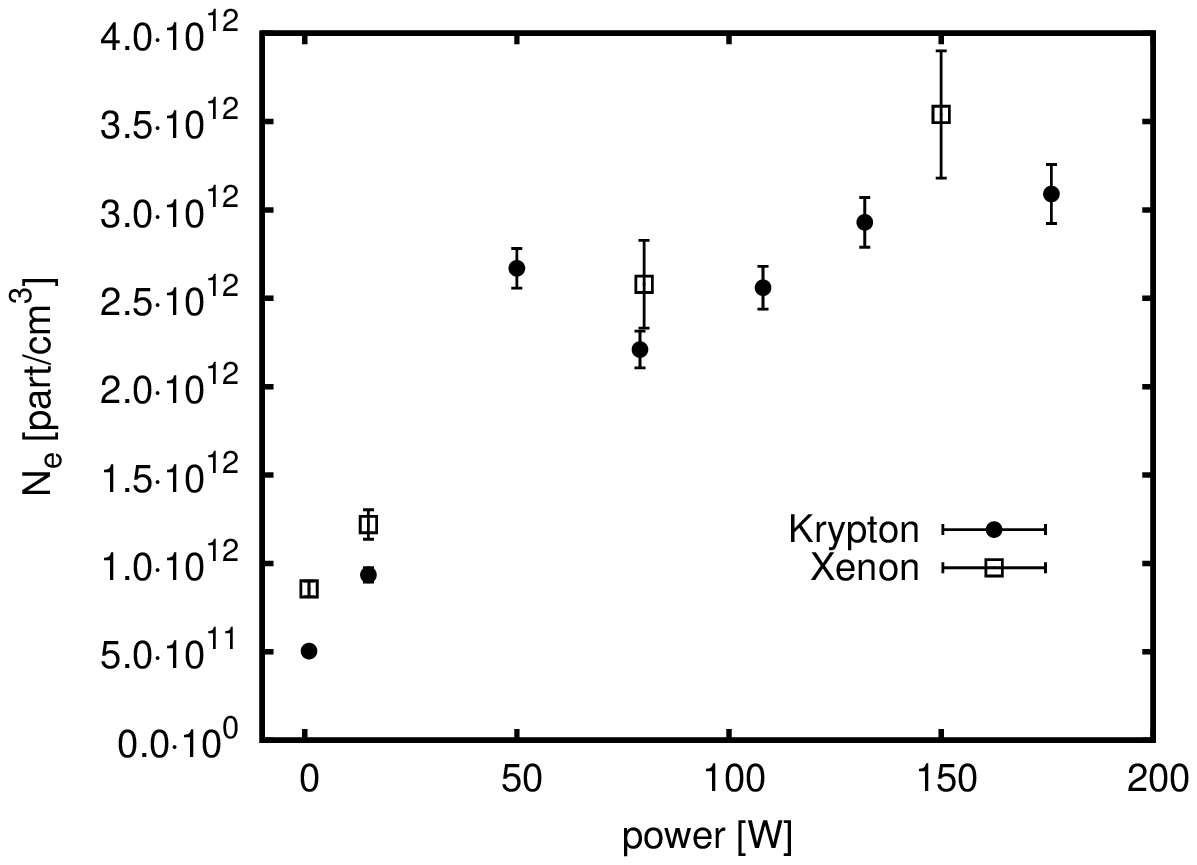}

\caption[]{Electronic density for Kr and Xe gases as a function of injected power.}
\label{fig:Ne}
\end{figure}

The values of $N_i$ and  $N_e$ obtained from the measured $\langle q \rangle$ and $N_i N_e$ product determined from the Bremsstrahlung spectra are presented in Figs.~\ref{fig:Ni} and \ref{fig:Ne} and Tables~\ref{tab:kr} and \ref{tab:xe}. 

As it can be observed, densities in the order of $10^{11}$ and $10^{12}$~part/cm$^3$ were observed, for ions and electrons, respectively, in agreement with previous measurements \cite{dkgb2000}. Each of the electronic and ionic density increases as well with the injected power.

\section{High-resolution X-ray spectroscopy of argon plasma}
\label{sec:xray}
\begin{figure}[htbp]
\centering
\includegraphics[width=\columnwidth,height=0.6\textheight]{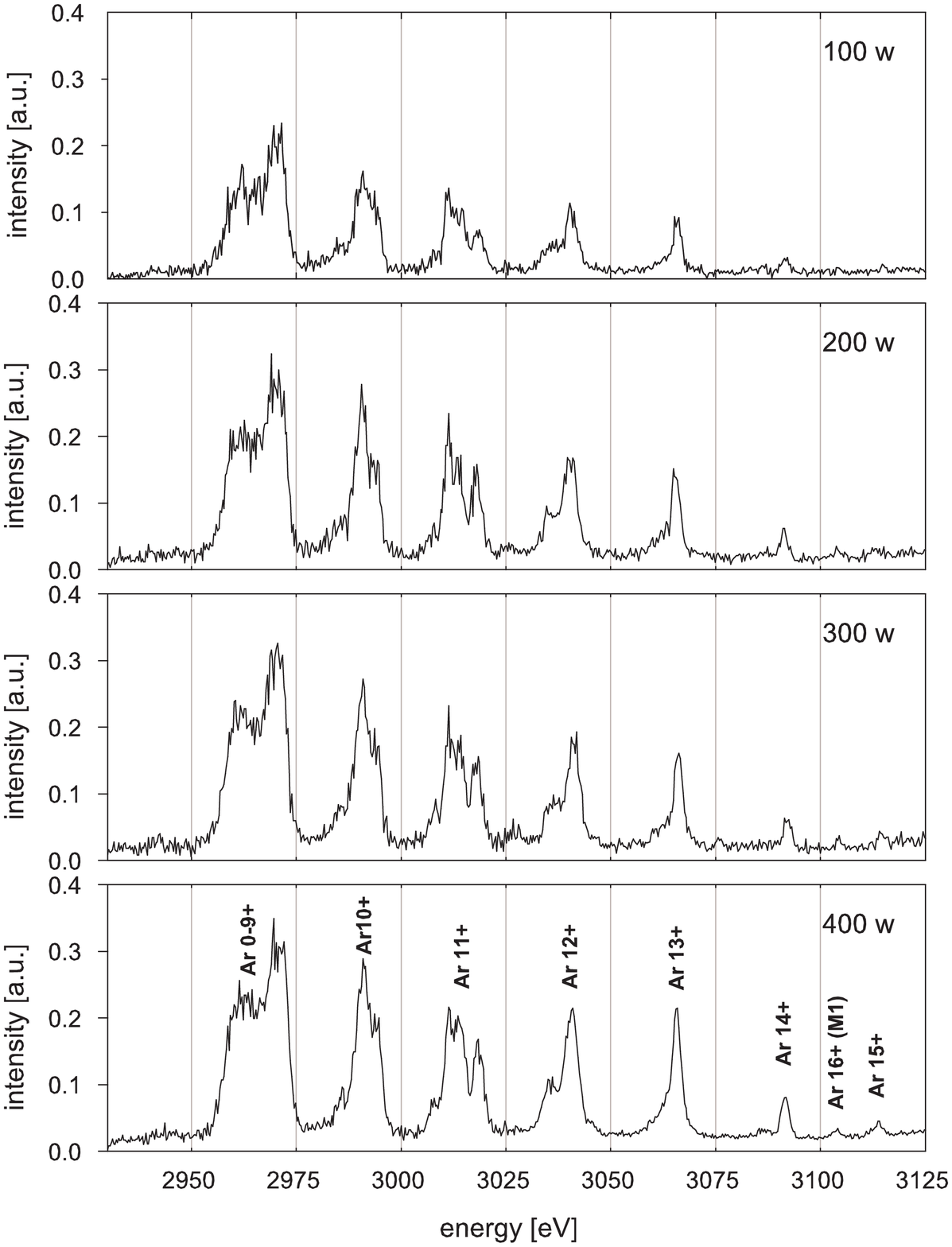}

\caption[]{Raw  X-ray spectra of Ar ions in the plasma, for several
microwave powers, measured with the mosaic-crystal X-ray spectrometer.
Each spectra have been normalized to the acquisition time.
}
\label{fig:ar-x-power}
\end{figure}

A very good way to understand the specific mechanisms that lead to the
different charge state  production and level populations of the different
ions is to observe the X-ray lines emitted by the plasma with a high resolution.
$K\alpha$ X rays from elements like argon provide a very good probe of their
ionization level in  plasma, because the electron undergoing a
transition to the $K$ shell for $q>8$ is in the same shell as the
electron removed by the ionization process. The energy shift
between different charge states are then relatively large. A typical good resolution
spectra from an Ar plasma thus displays an unresolved group of lines
containing $K\alpha$ lines from Ar$^{+}$ to Ar$^{9+}$ (i.e., ions with one
$K$ hole and 0 to 8 $M$ holes), and then well resolved features from
Ar$^{10+}$ to Ar$^{17+}$ (see Fig.~\ref{fig:ar-x-power}), depending on the conditions of the plasma.
The $K$ ($2p\to 1s$)
X-ray spectra of Ar from an ECR ion source was first observed using a cylindrically-bent
Johann-type X-ray spectrometer, with a 3.5~eV resolution ~\cite{dkgb2000}. A
detailed interpretation involving complete calculations with theoretical
energies and cross-section values has been published in Refs.~\cite{cmps2001,mcsi2001}.

Here we present good resolution ($2.5$~eV) spectra of argon plasma,
observed by the mosaic-graphite crystal spectrometer described in
Sec.~\ref{sec:mosaic-spectro}. 
The detector--crystal distance  ($L$, equal to the source--crystal one) 
has been chosen to cover an energy range from about 2900 to 3200~eV 
in order to record $K\alpha$ radiation from
 neutral to He-like argon. To account for very high count rates,
the spectrometer transmission has been reduced by a factor of 120, placing a 0.5 mm vertical slit (See~\ref{sec:mosaic-spectro})
in front of the graphite crystal. This results in a linearly varying total integrated
efficiency (from 1.8 to $2.7× 10^{-8}$) in the energy range relevant for Ar (from 2.8 to
3.2~keV). This evolution is mostly due to the variation of efficiency of the
position-sensitive proportional gas counter as function of photon energy.

The different spectra shown Fig.~\ref{fig:ar-x-power},  were recorded at 
injected microwave powers ranging from 100 to 400~W and have been normalized
to the total acquisition time and corrected for detector efficiency.
Those spectra were obtained with a copper plasma chamber, which should provide
plasma with lower electronic densities than the aluminum chamber 
used when  Bremsstrahlung spectra presented in
Sec.~\ref{sec:plasma} were studied. 
The typical acquisition time ranges from 1000 to 3000~s.
Oxygen was systematically used as a buffer gas,
and the source optimized to provide good extracted currents.
Here our main goal was to observe the variation of the spectra as a function of the injected power,
to identify the Ar lines, and to check whether the plasma of an ECRIS can be an X-ray source 
intense enough to perform high precision measurements of excited level energies of highly charged ions.

\begin{figure}[t]
\centering
\includegraphics[width=\columnwidth,angle=0]{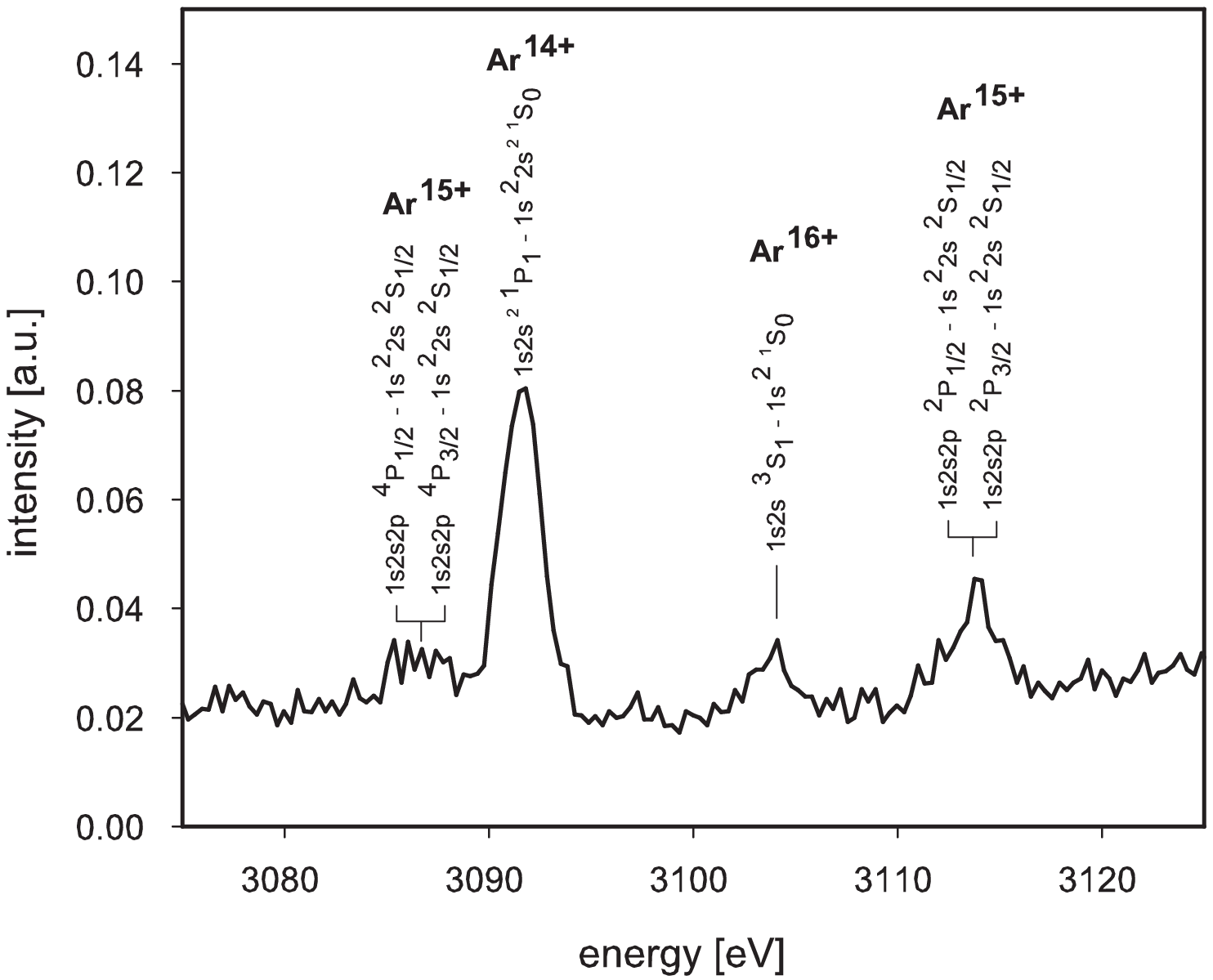}

\caption[]{ Zoom of the region around  3.1~keV recorded with the mosaic-crystal X-ray spectrometer for a microwave power of 400~W.
Identification of the Ar$^{14+}$ to Ar$^{16+}$ main transitions is presented.

}
\label{fig:ar-x-ident-line}
\end{figure}

It should be noted that the 2950--2975~eV line complex is present even at
1~W injected power.  At 400~W, the Ar$^{13+}$ is the prominent and would certainly be a good candidate
for a first attempt for high-precision X-ray spectroscopy measurements. We
also observed the $1s 2s^2 2p\, ^1P_1\to 1s^2 2s^2 \, ^1S_0$ Be-like
transition, the $1s 2s 2p \, ^2P_{3/2,1/2}\to 1s^2 2s \, ^2S_{1/2}$ Li-like transition
as well as the magnetic dipole (M1)  $1s 2s \,^3S_1\to 1s^2 \, ^1 S_0 $ He-like transition.
Identification
of the lines coming from charge states 14+ -- 16+ in the spectra obtained at 400~W is
presented on Fig.~\ref{fig:ar-x-ident-line}.

The detection of the relativistic M1 transition in He-like argon is particularly relevant.
Due to its small natural width (few $\mu$eV), this line would be the ideal candidate for high-precision,
 high-resolution spectroscopy,
allowing for accurate tests of Quantum Electrodynamics and relativistic effects in
highly charged ions,
but also for a new X-ray standard definition in the energy range of a few keVs \cite{agis2003}. Keeping in mind that the X-ray spectra, presented here, were recorded from a plasma not only collimated by a small 500~$\mu$m diameter pinhole but also with a vertical slit of 0.5 mm in order to reduce the counting rate, precise measurements using a very high-resolution spectrometer should be feasible.

\section{Conclusion}
\label{sec:concl}
In this paper we have performed a detailed analysis of the plasma of a
commercial, permanent-magnet ECRIS source with a mirror ratio of
$R=2.5$. We used the electron Bremsstrahlung as a tool to measure
electronic spectral temperatures.  We have shown that the temperature is weakly
dependent on the microwave power and has, in normal operating
conditions of the plasma source,  a value ranging from 39 to 50~keV, independently
of the injected gas. Even though the plasma volume of such a
permanent magnet ECRIS is necessarily small, the electronic
temperature of this source is four times larger than what has been found on a
10~GHz, $R=3.7$ source (T around 11~keV) \cite{dkgb2000}, and at the lower edge  of the range mesured on a 18~Ghz , $R=2.6$ ECRIS
(T between 35 and 100~keV) \cite{blbg1994}, as expected.  On the other hand,  the electronic and
ionic density product has been found to depend very strongly on the power and slightly on the injected gas.
In addition, we show how electronic and ionic densities can be estimated separately when measurements of extracted currents are combined to simultaneous records of Bremsstrahlung spectra. In agreement with previous measurements, the ionic density is in the range of $10^{11}$ and the electronic density in the range of $10^{12}$~part/cm$^3$, both increasing rapidly with the injected power. In parallel, a mean charge state of 8+ has been found for Kr ions and 9+ for Xe ions, but there independently of the injected power above 50W.

Finally,
we have demonstrated that the brightness of this small plasma ECRIS is sufficient to observe with a crystal spectrometer, in a short time, with a huge reduction in transmission, X-rays even from Ar$^16+$.
This  evidences for possible use of high-resolution low-efficiency spectrometers to perform high-precision measurements of transition energies of highly charged ions, especially since the running conditions can be rather easily improved by the implementation of a linear coupling of the microwave to the plasma and by increasing the microwave frequency as well.

\section*{Acknowledgments}
The SIMPA ECRIS has
been financed by a grant from CNRS, ``BQR equipment grant'' with a
``Plan Pluriformation'' and an
``Infrastructure'' grant from the ``Ministère de la Recherche et de
l'Enseignement Supérieur''. These experiments are  also partially supported by a
grant from ``Agence Nationale pour la Recherche (ANR)'' number
\emph{ANR-06-BLAN-0223} and Helmholtz Alliance HA216/EMMI. We thank J.P.~Okpisz, B. Delamour,
P.~Travers, A.~Vogt, C.~Rafaillac and S. Souramassing for technical
support in electronic and mechanics.
We wish to thank J.-Y. Pacquet and L. Maunoury for their help
when we obtained the first beam from SIMPA, and Pr. C.T. Chantler for
very enlightening discussions about X-ray absorption coefficients.

Laboratoire Kastler Brossel is ``Unité Mixte de Recherche du CNRS, de
l'ENS et de l'UPMC n$^{\circ}$ 8552''. L'Institut des Nanosciences de
Paris (INSP) is ``Unité Mixte de Recherche du CNRS et de l'UPMC n$^{\circ}$ 7588''.


\end{document}